\documentclass[runningheads]{llncs}
\usepackage{graphicx}
\usepackage[table]{xcolor}
\usepackage{graphicx}
\usepackage{hyperref}
\begin{document}
\title{Sound effects in media:A comparative analysis of recorded and synthetic samples in live-action and animation.}
\titlerunning{Sound effects in media: A comparative analysis.}
\author{Nelly Garcia \inst{1} \orcidID{0009-0007-3986-4518}\and
Joshua Reiss\inst{1}}
\authorrunning{Garcia N. et al.}
\institute{Centre for Digital Music (C4DM),Queen Mary University of London, 327 Mile End Rd, Bethnal Green, London E1 4NS 
\email{n.v.a.garcia-sihuay@qmul.ac.uk}\\ \doi{0009-0007-3986-4518} 
\email{joshua.reiss@qmul.ac.uk}
\url{https://www.qmul.ac.uk/}}
\maketitle              % typeset the header of the contribution
\begin{abstract}
Creating sound for storytelling is crucial to establishing the environment in productions such as films, TV series and video games. This process often involves repeating, layering and recording real objects or using sound libraries, which can be time-consuming and repetitive. To address these challenges, procedural audio, also known as digital foley, offers a solution by allowing sound designers to quickly generate samples. Despite its efficiency, questions remain about the believability of synthetic samples compared to real ones. In our study, we compared synthetic samples generated by an online procedural engine and integrated them with both animated and live-action visuals. Our results indicate that procedural audio is highly effective and perceived as believable in drama and sci-fi scenes, particularly for sound models such as lasers, hits, air and rockets, whereas synthetic sounds weren't as believable in cartoon productions when representing everyday actions. Finally, we identified specific models that needed optimisation and highlighted audio features that needed improvement with feedback from audio professionals.

\end{abstract}
\section{Introduction}
 Sound has two basic roles in film, tv or games. These are story telling and story supporting. Sound is usually exaggerated (´hyper realism´) to create some reaction to the audience. \cite{dakic}, refers to story telling as the most important characteristic of films by using dialogues, monologues or off-narration. While story supporting relates on the effects that inhence the tension in film and suggest the audience how to feel.  \cite{low}, states that sound design was properly introduced as a term in a Francis Coppola's movie. \cite{susini}, defines sound design as an approach to creating 'new' sounds with the intention that they will be heard in a given context of use.
 
 Sound designers' workflows vary depending on the sources they are working with, such as libraries, Foley, direct sound and re-recording. Foley, named after Jack Foley, \cite{foleygrail} refers to creating the sound of the object using different objects. (e.g. coconut shells for the horse steps).

Sound designers need to consider the period and genre of a story, as well as the venue where it takes place, previous work had interviewed professionals and stated their process \cite{pauletto}.They need to respond to the social, historical and cultural context of the production.This context requires deciding which sounds to include in the soundscape. \cite{Rochesso}. Finding the right sound for an object can be time consuming and involve listening to different samples repeatedly. The hypothesis that creating sounds without spending minutes searching for the right one can help sound designers innovate their libraries, the process of sound designed explain as a scientific method \cite{ircam} has confirmed this issue. For example, the sound of a monster can be created from scratch, offering a wide range of models simultaneously, rather than layering sounds. Procedural audio engines give users a wide range of parameters to modify sounds.

Though the perception of the "believability" of sound design could be subjective and depends on which story the sound designer wants to share. \cite{AQA}, states a division of the different ways a sound design could be created: 
\begin{itemize}
    \item \textbf{\textit{a. Realistic:}} Incorporation of elements that are meant to represent real life. It's important to mention the difference between \textit{realism} and \textit{reality}. \cite{realism}, described realism as ‘the illusion of reality’,as a representation, without limiting it to truthfulness or authenticity. Reality on the other hand is real life as we experience it.
    \item \textbf{\textit{b. Symbolist:}}Interest in communicating an idea to the audience rather than representing real life. E.g a whirling sound when the scene transports you to the mind of the character. 
    \item \textbf{\textit{c. Fantasy:}} Allowance to create a new world. Use of a wider range of sound effects,music or soundscapes to create a magical atmosphere.
\end{itemize}

A concept related to the perception of reality used nowadays is \textit{hyperrealism}, \cite{hyper}, which means exaggeration of the image of reality, also understood as making the sound bigger than it is perceived in reality. Overall, soundscapes set location or create a mood. Most designers will use a combination of these types of sound to create the design for a production. According to \cite{Lopez} the perceived soundscape of a sound effect depends on two key factors:

\begin{enumerate}
    \item \textbf{The director's perspective:} The director's vision and storytelling are paramount. Directors often describe the desired environment to the sound designer, providing a visual and conceptual framework for how the sound should be perceived.
    
    \item \textbf{Audience perception:} The effectiveness of a sound effect is ultimately determined by the audience's perception. The goal is to evoke the intended emotional response and enhance the overall viewing experience.
\end{enumerate}

Once the story and the target audience is decided, we need to discuss the possibilities for creating the audio files that will be in the production.We considered sound synthesis as the artificial creation of sound using analogue, digital or a combination of both \cite{MoffatSS}. Procedural audio, also known as digital foley, uses algorithms to generate audio in real time, adapting to changing inputs. Also, procedural audio involves building signal synthesis and processing chains that respond to on-screen events. However, the complexity of implementing DSP algorithms for artistic tasks, combined with the lack of evaluation methods and comparisons, has made this approach less common in the industry, leading to a preference for sample-oriented methods in sound design. Analysing and understanding the characteristics of sounds in both sound synthesis and procedural audio can improve algorithms for different sound categories, establish correlations between algorithms and perceptual concepts to provide objective explainability of the computer processes. This approach provides valuable insights for improvement e.g. \cite{Whoosh}, did an approach trying to describe the 'whoosh' sound and how it could be improved pointing out the different audio descriptors needed to create form scratch a sound of this type.

This project aims to bridge the gap between research and industry practice by comparing the 'realism' and 'believability' of synthesis methods with traditional library samples already integrated with visuals. Our research questions are: 
\begin{itemize}
    \item \textit{How believable are synthetic sounds compared to traditional library samples when combined with visuals?}
    \item \textit{In which scenarios do procedural audio models perform best in terms of audience perception?}
    \item \textit{Which specific models and audio features in procedural audio need further optimisation?}
\end{itemize}
In this paper we discuss the differences in the perception of realism between synthetic and real(library) audio samples. Section 1.1 introduces the methods and previous work in the field. In Section 2, we outline the experimental design and the dataset used. Sections 3 and 4 deal with the qualitative and quantitative analysis of our results. Finally, Section 5 presents our conclusions.

\subsection{Methods}
To answer the first and third research questions, our primary goal is to identify low-level audio features that can help audio developers and sound designers improve algorithms for creating sounds from scratch. Audio features are defined as contextual information that can be extracted from an audio signal,they are a statistical or computational representation of an audio file. According to \cite{ObjMoffat}, there could be a generalisable objective set of measures for synthesised sounds. His work provided a comparative table of different audio descriptors commonly used in the evaluation of synthetic models which provided interpretability for sound improvement. We define the interpretability as the capability of understanding how the model works, i.e., the process that led to its output \cite{interp}.

As mention in section 1. procedural audio can be a key aspect in the sound design process, it explores the creativity and has more areas of opportunity:

\begin{itemize}
    \item \textit{Optimise the design process:} Giving sound designers greater control over procedural audio algorithms can help them to produce better samples more efficiently.
    \item \textit{Improve procedural audio algorithms:} The identified audio characteristics can guide the creation of sound families that closely match the characteristics of real sounds.
    \item \textit{Cost-effective sound creation:} Procedural audio models can generate a wide variety of sounds, reducing the need for extensive studio time.
\end{itemize}

\begin{figure}
    \centering
    \includegraphics[width=1\linewidth]{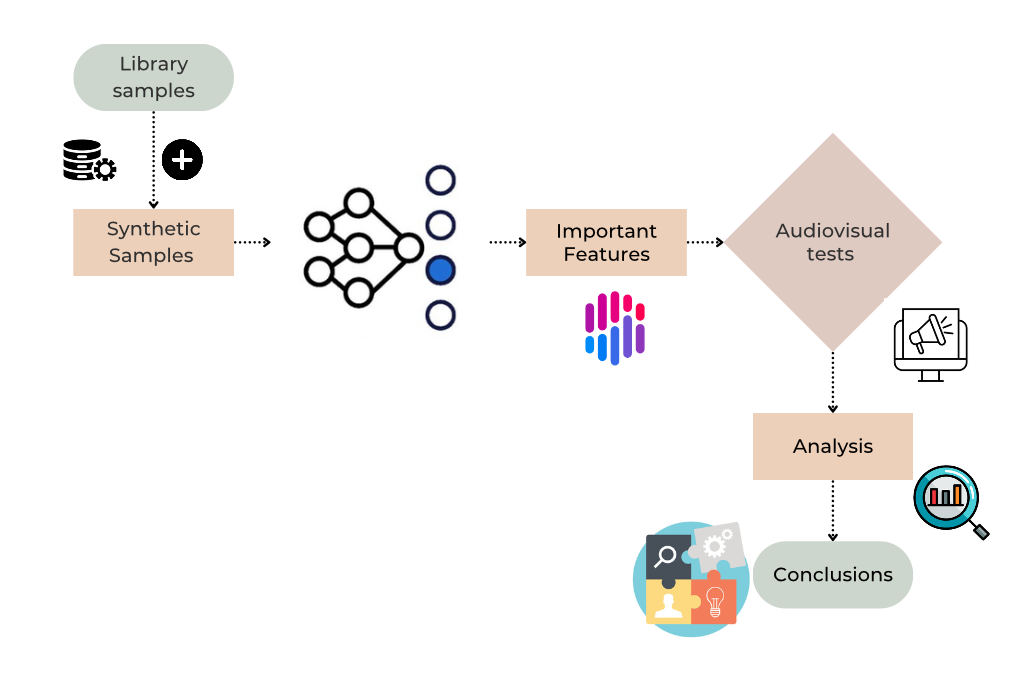}
    \caption{Process followed to compare the objective and subjective audio descriptors}
    \label{process}
\end{figure}

Through our study, we discovered several features that provide insight into specific aspects of sound, allowing sound designers to make relevant improvements.The process of this methodology is shown in figure \ref{process}.

The first step in the process shown above, was to create a dataset of different videos intended for sound design, which is described in detail in Section 2. We normalised and extracted audio features from these videos, comparing each sound category using a bi-classification machine learning model.

We used XGBoost \cite{xgboost} and Random Forest \cite{rfc} models to tackle the bi-classification problem, achieving 95\% and 90\% accuracy respectively. Once the models produced results, we applied SHAP (SHapley Additive exPlanations) \cite{shap} and PCA (Principal Component Analysis) \cite{PCA} to interpret the results. We identified the top four audio descriptors for each sound category and conducted subjective tests to determine whether these audio features made a noticeable difference to the audience.

For the analysis of the audio features, we used the state-of-the-art reference \cite{cuidado} to classify and distinguish the retrieved audio features in different sound categories.

\begin{itemize}
    \item \textit{Steadiness or dynamics: }Features represent values extracted from the signal at a given time, such as mean and standard deviation.
    \item \textit{Temporal extent:} Descriptions apply to specific parts of the sound, such as attack and loudness.
    \item \textit{Abstractness: }This refers to what the feature represents, such as cepstrum and linear prediction.
    \item \textit{Extraction process:} Depending on how the data was extracted, features can be
    \begin{itemize}
        \item Calculated directly on the waveform data.
        \item Extracted after performing a signal transform (e.g., Zero Crossing Rate, FFT, Wavelet).
        \item Related to the signal model (e.g., source/filter model).
        \item Designed to mimic the output of the auditory system (e.g., Bark or ERB bank filter output).
    \end{itemize}
\end{itemize}

The analysis of the subjective and objective ratings is discussed in sections 5 and 6.

\section{Dataset and extraction}

\begin{figure}
    \centering
    \includegraphics[width=1.0\linewidth]{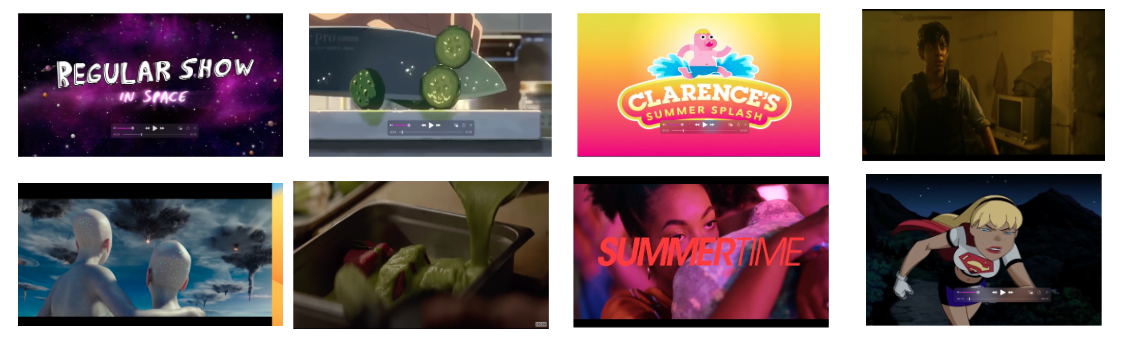}
    \caption{Scenes for the visual evaluation}
    \label{scenes}
\end{figure}

 We selected eight scenes, split evenly between live action and animated formats. These scenes, can be found in YouTube \footnote{https://www.youtube.com/@nellyngz/videos}, had all their original sounds replaced with custom sound designs created specifically for each scene. Each video, which ranged in length from 30 to 60 seconds, covered a range of genres including action, drama, sci-fi and comedy. We used between 50 and 60 different sounds for each sound design. Figure \ref{scenes} illustrates the selected scenes, with the cartoon scenes at the top and the live action scenes at the bottom.

For sound samples we accessed the BBC Sound Effects library\footnote{\url{https://sound-effects.bbcrewind.co.uk}}, the Hybrid Sound Library, 50-ESC dataset \cite{pic} and the Soundsnap online library\footnote{\url{https://www.soundsnap.com}}. For the synthesis samples we used Nemisindo\footnote{\url{https://nemisindo.com}}, an online procedural audio engine. Each video had specific sound requirements (e.g. for the sci-fi videos we needed spaceship movements and lasers). Table 1. Shows the different synthetic models that were used and combined in the sound design for evaluation. The synthetic models retrieved from the procedural audio engine are shown in the third column. Approximately 16 synthetic sound models were used in all sound designs. Some synthetic models were modified to create different sounds (e.g. the explosion model was used to create a splash).

The scenes contained  different sound synthesis models. Each sound design consisted of between 30-40 sounds between real and synthetic samples. Table \ref{scenes} shows which and how many synthetic models were used in each sound design.

\begin{table*}[ht]
\centering
\begin{tabular}{| m{3cm} | m{3.5cm} | m{5cm}|}
\hline
\rowcolor{lightgray}\textbf{Scene} & \textbf{Type-Genre} & \textbf{Synthetic Sounds} \\ \hline
\textit{Space War} & Sci-fi Live-action & Lasers, whoosh, rocket, drill, mammal sound, bells, spaceship movements \\ \hline
\textit{Space Adventure} & Sci-fi Cartoon & Lasers, rocket, drill, mammal sound, bells, machinery, increasing beams, space cannons \\ \hline
\textit{Rescue Mission} & Action Live-action & Gunshot, hits, impact, debris, body hits, explosions, clothes movements \\ \hline
\textit{Superhero Fight} & Action Cartoon & Fire embers, crackling, hits, impact, debris, body hits, explosions, clothes movements \\ \hline
\textit{Ramen} & Drama Cartoon & Hot water, whoosh, droplets, fire embers, machinery \\ \hline
\textit{Kitchen} & Drama Live-action & Hot water, whoosh, droplets, fire embers, machinery \\ \hline
\textit{Vacations} & Drama/Comedy Cartoon & Splash, waves, hit, air, birds, whoosh, spray, bubbles, droplets \\ \hline
\textit{Summertime} & Drama/Comedy Live-action & Splash, waves, hit, air, birds, body hit, whoosh, spray, bubbles, droplets \\ \hline
\end{tabular}
\caption{Scenes used in the evaluation of the sound design and synthetic sounds used in the video.}
\end{table*}

Each scene had three types of sound design, which are listed below:
\begin{itemize}
\item \textbf{Real sound design:} This sound design contained only library samples.
\item \textbf{Synthetic sound design:} This sound design contains a mix of library and synthetic samples.
\item \textbf{Bad sound design:} This sound design contained random sounds that were not synchronised with the image. The purpose of this sound design is to filter the participants, to give them the context of the story being revealed in the scene.
\end{itemize}

\section{Sound designs  objective analysis}
For its analysis, we created a dataset of 50 samples for each sound category and compared them to the library samples we had access to. The dataset consists of a total of 1,616 sounds divided into 32 categories (16 synthetic and 16 library samples). Each category varies from 2.5\% to 2.8\% to achieve balance.

The costumed dataset previously built with the procedural online engine and the libraries mentioned above in section 3. All of our samples for analysis were downmixed to mono, 16-bit depth, 44.1 Khz and trimmed to 5 seconds.

Following the criteria as the \cite{cuidado} project and \cite{ObjMoffat}, we determined the type of audio descriptors to be extracted. We used the Essentia framework \footnote{https://essentia.upf.edu} \cite{essentia} for feature extraction, with a total of 78 low-level features for each sound sample. These descriptors were carefully selected and combined to extract the most critical features that allow the machine to distinguish between synthetic and realistic sounds.

After evaluating the performance of the classification model, we identified the four best audio descriptors for each sound category. Table 2 shows the top audio descriptors for each sound category, divided by the feature domain where each audio descriptor can be found. The last column shows the general classification of the feature, as explained earlier in this section.

These descriptors help to understand how synthetic sounds produced by models can be compared to physical sounds. Audio descriptors belonging to the temporal and frequency domains are highlighted in green, indicating that changes in these domains could improve the perception of these synthesis models. Descriptors highlighted in yellow belong to the temporal and spectral domains, while those highlighted in purple are purely temporal features.
\begin{enumerate}
    \item \textit{Temporal shape:} features(global or instantaneous),computed from the waveform or the signal envelope(envelop).
    \item \textit{Spectral features:} computed by the STFT,centroid, spread, skewness, roll off frequency
\end{enumerate}
\begin{table*}[ht]
\begin{tabular}{||p{2.5cm}|p{4.5cm}|p{3.5cm}|p{3cm}||}
  \hline
  \rowcolor{lightgray}\textbf{Category} & \textbf{Audio Features} & \textbf{Feature Domain} & \textbf{Classification} \\
  \hline
  \textit{Fire} & High Frequency Effective Duration Flatness SFX & \cellcolor{green!25}Temporal and Frequency & Extraction and Time Extent \\
  \hline
  \textit{Rain} & TCToTotal, Effective Duration, Kurtosis, Flatness SFX & \cellcolor{blue!25}Temporal & Dynamicity and Time Extent \\
  \hline
  \textit{Waves} & TCToTotal, Log Attack Time, Flatness SFX, Strong Peak & \cellcolor{green!25}Temporal and Frequency & Extraction and Time Extent \\
  \hline
  \textit{Wind} & Log Attack Time, TCToTotal, MaxToTotal, Pitch Salience & \cellcolor{blue!25} Temporal & Extraction and Time Extent \\
  \hline
  \textit{Engine} & Pitch Salience, Attack Start, TCToTotal, Flatness SFX & \cellcolor{green!25}Temporal and Frequency & Extraction and Time Extent \\
  \hline
  \textit{Applause} & Intensity, TCToTotal, MaxToTotal, Pitch Salience & \cellcolor{green!25}Temporal and Frequency & Extraction and Dynamicity \\
  \hline
  \textit{Gunshot} & TCToTotal, MaxToTotal, ZCR, High Frequency & \cellcolor{green!25}Temporal and Frequency & Extraction and Dynamicity \\
  \hline
  \textit{Helicopter} & TCToTotal, Intensity, Effective Duration, High Frequency & \cellcolor{green!25}Temporal and Frequency & Dynamicity and Time Extent \\
  \hline
  \textit{Thunder} & Roll Off, TCToTotal, ZCR, MaxToTotal & \cellcolor{yellow!25}Temporal and Spectral & Extraction and Time Extent \\
  \hline
  \textit{Crackling} & Log Attack Time, TCToTotal, Attack Stop, Effective Duration & \cellcolor{blue!25} Temporal & Time Extent \\
  \hline
  \textit{Fireworks} & ZCR, Roll Off, TCToTotal, Kurtosis & \cellcolor{blue!25}Temporal & Extraction and Time Extent \\
  \hline
  \textit{Bells} & TCToTotal, Pitch Salience, MaxToTotal, ZCR & \cellcolor{blue!25}Temporal & Extraction and Time Extent \\
  \hline
  \textit{Pouring Hot Water} & TCToTotal, MaxToTotal, Roll Off, ZCR & \cellcolor{yellow!25}Temporal and Spectral & Extraction and Time Extent \\
  \hline
  \textit{Explosion} & TCToTotal, MaxToTotal, High Frequency, Roll Off & \cellcolor{yellow!25}Temporal and Spectral & Extraction and Time Extent \\
  \hline
  \textit{Bubbles} & ZCR, TCToTotal, Flatness, MaxToTotal & \cellcolor{blue!25}Temporal  & Extraction and Time Extent \\
  \hline
\end{tabular}
\caption{Top 4 Audio Features Retrieved by Machine Learning Methods and SHAP Values}
\end{table*}

The previous listed sound descriptors were extracted with Essentia. To make more sense in how this audio features were extracted, the meaning of each top feature are listed below:

\begin{itemize}
    \item \textbf{TCToTotal}:This algorithm calculates the ratio of the temporal centroid to the total length of a signal envelope, indicating how 'balanced' the sound is. Its value is close to 0 if most of the energy is at the beginning of the sound (e.g. decrescendo or impulsive sounds), close to 0.5 if the sound is symmetrical (e.g. 'delta unvarying' sounds), and close to 1 if most of the energy is at the end of the sound (e.g. crescendo sounds).
    \item \textbf{High Frequency}:  Content of a spectrum, The magnitudes of the spectral bins are added together, but multiplying each magnitude by the bin "position" (proportional to the frequency). Thus if X(k) is a discrete spectrum with N unique points, its high frequency content measure is represented in equation 1:
    \begin{equation}
       HFC= |X(n)|^2 \cdot k\label{eq:your_equation}
  \end{equation}
    \item \textbf{Flatness SFX}:There are two thresholds defined: a lower one at 20 percent and an upper one at 95\%. The thresholds yield two values: one value which has 20\% of the total values underneath, and one value which has 95\% of the total values underneath. The flatness coefficient is then calculated as the ratio of these two values.
    \item \textbf{Effective Duration}: Effective duration of an envelope signal. The effective duration is a measure of the time the signal is perceptually meaningful. This is approximated by the time the envelope is above or equal to a given threshold and is above the -90dB noise floor. This measure allows to distinguish percussive sounds from sustained sounds but depends on the signal length.
    \item \textbf{Kurtosis}:Measure of the "tailedness" of the probability distribution of a real-valued random variable. Describes a particular aspect of a probability distribution.
    \item \textbf{LogAttackTime}:Calculates the logarithm (base 10) of the attack time of a signal envelope. The attack time is the time from when the sound becomes perceptually audible to when it reaches its maximum intensity. By default, the start of the attack is estimated at the point where the signal envelope has reached 20\% of its maximum value, to account for possible noise, and the end is estimated at the point where the envelope has reached 90\% of its maximum value, as in trumpet sounds.
    \item \textbf{PitchSalience}: This algorithm computes the pitch salience of a spectrum. The pitch salience is given by the ratio of the highest auto correlation value of the spectrum to the non-shifted auto correlation value. Pitch salience was designed as quick measure of tone sensation.
    \item \textbf{Zero Crossing Rate:} It is the number of sign changes between consecutive signal values divided by the total number of values. Noisy signals tend to have higher zero-crossing rate.
    \item \textbf{MaxToTotal: } Computes the ratio between the index of the maximum value of the envelope of a signal and the total length of the envelope. This ratio shows how much the maximum amplitude is off-center.
\end{itemize}

\section{Experiment Design}
\begin{figure*}
    \centering
    \includegraphics[width=0.6\linewidth]{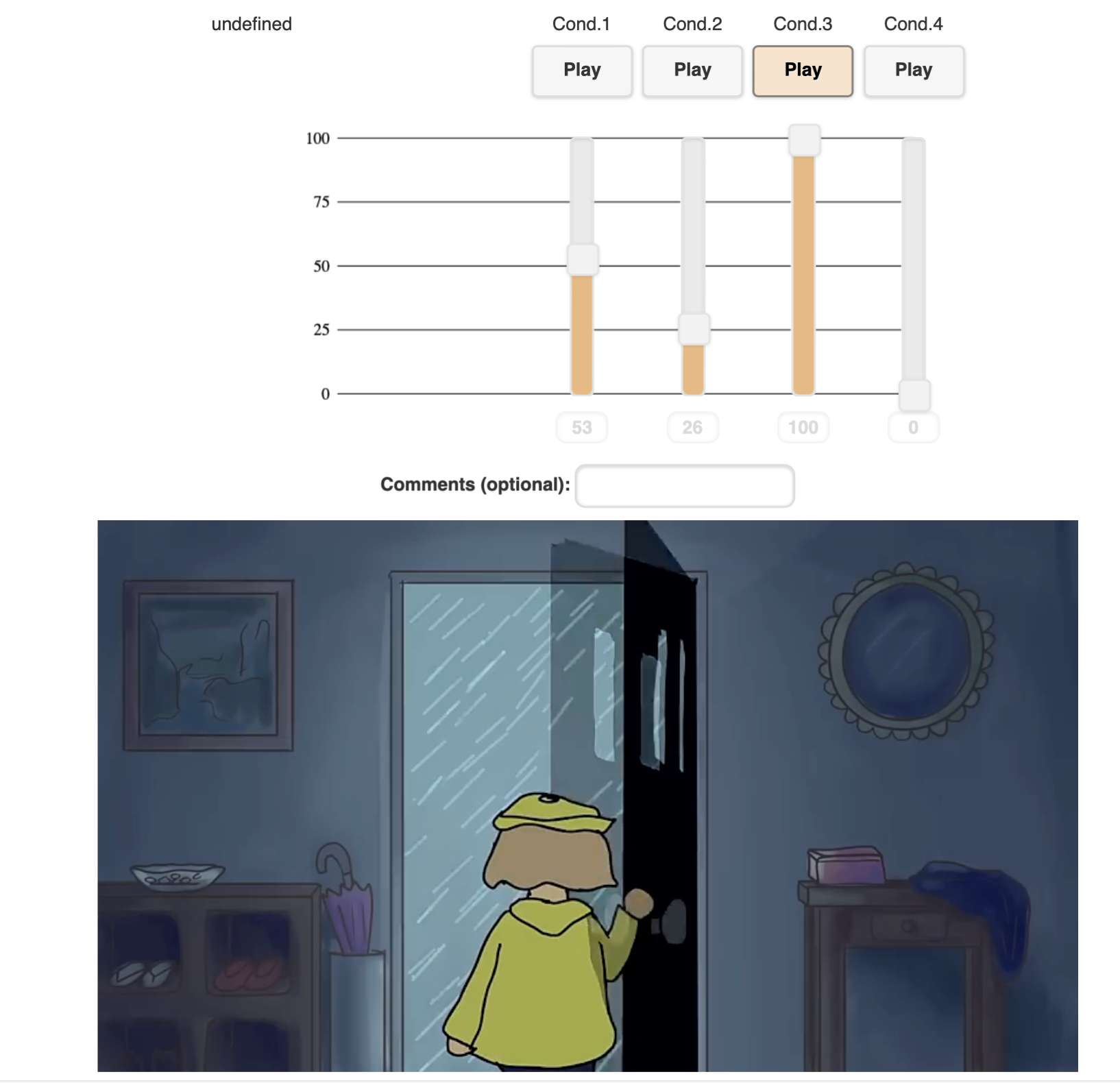}
    \caption{WebMUSHRA Test with no reference and three types of sound design}
    \label{webmush}
\end{figure*}

Our subjective evaluation consisted of two parts. The first part was to evaluate the perception of the final videos and the second part was finding which sound effects weren't perceived  as believable and need to be optimized. We created a MUSHRA-inspired test using WebMUSHRA \cite{webmushra}, 4 sliders were used to rate the different sound designs as shown in shown in figure \ref{webmush}. Our rating was from 0-100 with the following scale:  

\begin{itemize}
    \item \textbf{100 - Good sound design. }(The samples help to create a real soundscape.)
    \item \textbf{75 - Mostly good sound design.} Not that real(The samples do not help to create a real soundscape.)
    \item \textbf{50 - Average sound design} (The samples used are mostly synthetic, they do not help to create a real soundscape.)
    \item \textbf{25 - Somewhat bad sound design }(The samples do not help to create a real soundscape.)
    \item \textbf{0 - Bad sound design }(The samples are not coherent with the visuals.)
\end{itemize}

And for the second part of the evaluation, we designed the Google form \footnote{\url{https://docs.google.com/forms/d/e/1FAIpQLSd4_IwgM0plWo2ug5Odu89mgm3yYfWCrwwrR1e75-iryGI3aw/viewform}} where participants could indicate which sounds did not seem believable and suggest aspects that needed to be changed or could improve perception.

\section{Quantitative Evaluation}
For the webmushra quantitative evaluation. We had 20 participants, 11 male and 9 female, who performed the online test. All participants had from 3-7 years of experience in the audio industry, Overall, they are sound designers(40\%), audio researchers(30\%), developers(15\%) and musicians(15\%). The answers of two participants were discarded, because the rating for bad sound design was $\geq25$.

With the 18 remaining responses, we performed an ANOVA test, the \textit{p} and \textit{f} values are presented in the table \ref{ANOVAmushra} below, highlighted in bold are the sound designs' p-value less than 0.05. We found statistical importance difference in 5 out of the 8 sound designs. Two of these sound designs are cartoon visuals and three are live action scenes. At the same time, we performed the Bonferroni correction\cite{bonferroni} to reduce the probability of making errors, following (($\alpha$= 0.05/8)), where the numerator represents the value of alpha and the denominator is the number of cases, this lead to a change of alpha to \textit{0.00625 }to better understand the statistical difference. With this correction we found an statistical difference in 2 main scenes that represented daily routine sounds (Drama-kitchen scene) and the sci-fi cartoon scene where more imaginative sounds were used. As, mentioned in section 1, realism is defined as the representation of life without limiting it to truthfulness or authenticity.  This results point out that even though synthetic sound could be used easily in cartoons, if the cartoons mimic a daily routine, audiences tend to spot the synthetic sound easily,whereas in sci-fi, the hyperrealism or tendency to make a sound bigger could point out why synthetic samples weren't perceived as believable.

\begin{table*}
    \centering
    \begin{tabular}{| m{3.5cm} | m{2cm} | m{2cm} |m{2cm}|}
      \hline
       \rowcolor{lightgray}\textbf{ Scene} & \textbf{p-value} & \textbf{f-value} &\textbf{Bonferroni correction} $\alpha =0.00625$  \\
       \hline
        Action (LA) & \textbf{0.00283}  & 1.408 & 0.02  \\
        \hline
        Action (C) & 0.015 & 9.14 &0.12\\
        \hline
        Drama (LA) & \textbf{0.000092} & 6.05& \textbf{0.0007} \\
        \hline
        Drama (C) & \textbf{0.0000014}3 & 7.5& \textbf{0.00001 }\\
        \hline
        Drama/Comedy (LA) & 0.0016 & 1.66 &0.013\\
        \hline
        Drama/Comedy (C) & 0.114 & 8.66&0.919 \\
        \hline
        Space battle (LA) & \textbf{0.0017} & 5.3& 0.013 \\
        \hline
        Space (C) & \textbf{0.0001} & 5.4 & \textbf{0.0008}\\
        \hline
    \end{tabular}
    \caption{p and f values of the ANOVA  with the Bonferroni correction with alpha=0.00625 tests performed on the webmushra ratings. LA: Live-action. C: Cartoon}
    \label{ANOVAmushra}
\end{table*}

\begin{figure*}
    \centering
    \includegraphics[width=1.0\linewidth]{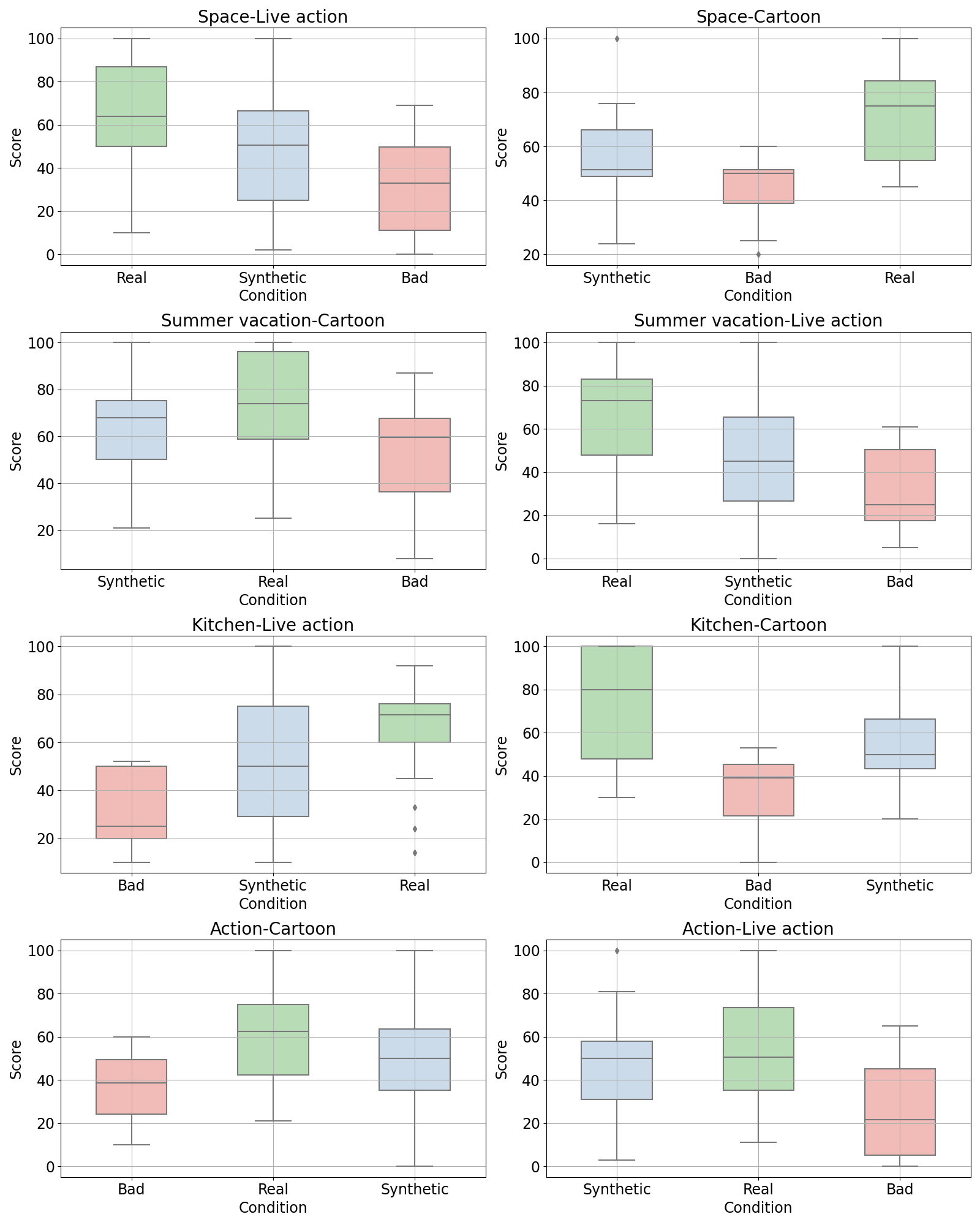}
    \caption{Boxplot of the 3 different sound designs in the different scenes.}
    \label{boxplotmushra}
\end{figure*}

For a better visualisation of this analysis, figure \ref{boxplotmushra} represents the different rating in the corresponding scenarios. The x-axis labels represents  the different sound designs used in the test, while the y-axis shows the ratings (0-100). It is easy to see that real sound designs were perceived as more believable for both cartoon and live action. Although the mean averages in cases like the summer holiday cartoon and the action live-action scenes aren't far apart. Performing the confidence levels we found out the maximum of 34.8\% of difference in the summer vacation scene between the real and the synthetic, and the least difference was found in action genre in live action with 13.8\%. An important observation is that the participants in the cartoon scenes rated the scenes between 50-70, which means that the models needed to be optimised or that people subconsciously think that cartoon sounds are not real because they don't associate them with a real object. 

\section{Qualitative Evaluation}

To understand more the results shows in figure \ref{boxplotmushra} we performed a more qualitative form of evaluation. Two participants' responses were discarded because they said they did not remember the sound effects at all and not feel comfortable taking them into account. 

\begin{figure*}
    \centering
    \includegraphics[width=1.0\linewidth]{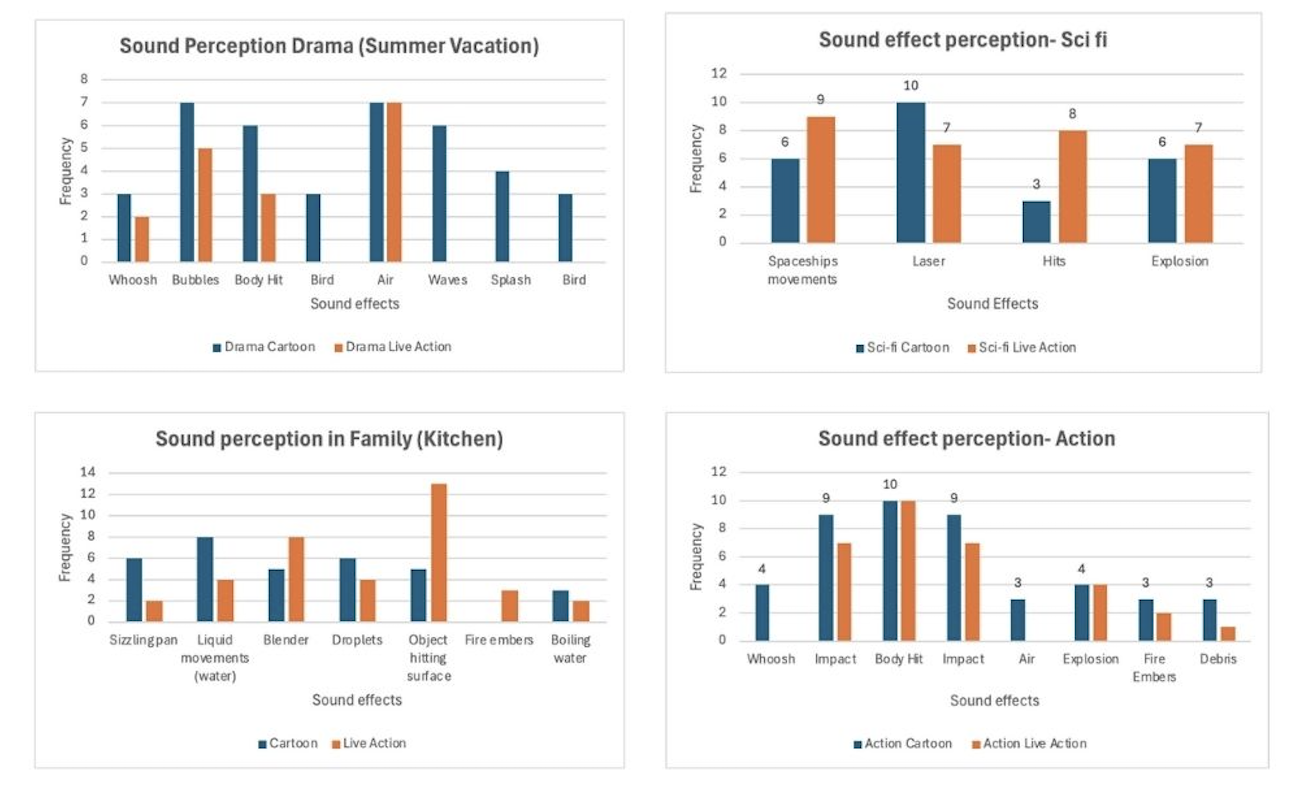}
    \caption{Barplots of the different sound suggestions of the participants}
    \label{barsound}
\end{figure*}

Participants had to choose 2 models that they did not find realistic in the sound designs, followed by a multiple choice question about which aspect could be optimised to make the sound more believable. 
Figure\ref{barsound} shows the different sounds chosen by the participants. The orange bar represents the live action, while the blue bar represents the animated scenes. We can see that there was less choice in the sci-fi (space) scene, where only 4 sound effects were seen. We can also see that although the laser was less credible in the animation, this doesn't apply to the live-action use, where the spaceship's movements were the most selected feature. 
However, there was more agreement that the cartoon sounds did not quite hit the mark, one hypothesis being that they weren't big enough to achieve the hyper-realism that the audiovisual project could have had. 
While more options were selected in the drama and action scenes, it was more obvious that the synthetic samples were easier to identify in the cartoon, leading to the hypothesis that the object shown in the visuals and the sound effect did not manage to get the "added value". 

We can see that the choice of sound supports the results of the webMUSHRA test and that there is a difference between cartoon and live action. We performed basic statistics such as mean, median, standard deviation, confidence levels,etc., and found variability in the participants' scores. The scenes with the most selected sounds were the drama/comedy and action scenes, each of which had 8 selected sound models, followed by the drama scene and finally the sci-fi scene. Our results did not come as a surprise, as people tend to match what they see with what they hear, according to the "added value" that sound can give to an image. For example, sci-fi sounds are mostly synthetic, but the hyper-realism of Hollywood productions means that sci-fi sounds require more post-production effects. Also, when it comes to drama or action scenes, people already have an idea of how certain sounds might be heard. 

However, these tests help us to highlight other aspects that have not yet been explored in the field of sound effects. Because of the lack of pure sound effects creations or tracks without music. When asked which aspects could be changed in their models, the participants selected them. We correlated the aspects with the audio features mentioned in section 2. Table 4 shows the most selected synthesis models in each genre, highlighting the participant's opinion in which aspect the model needs to be optimised and the feature domain to which this aspect correlates and which can be found in Table 2. 

\begin{table*}[h]
\centering
\resizebox{\textwidth}{!}{
    \begin{tabular}{| p{2.5cm} | p{3.8cm} | p{3.8cm} | p{3.5cm} |}
      \hline
       \rowcolor{lightgray}\textbf{Scene} & \textbf{Sound} & \textbf{Aspect} & \textbf{Feature domain} \\
       \hline
        \textit{Action (LA)} & 
        Body hit\newline Impact\newline Explosion & 
        More low frequencies\newline More low frequencies\newline Pitch modulation & 
        Frequency \\
        \hline
       \textit{Action (C)} &  
       Body hit\newline Impact\newline Debris & 
       More low frequencies\newline More low frequencies\newline Pitch modulation & 
       Frequency \\
        \hline
      \textit{Drama (LA)} & 
      Object hit\newline Blender\newline Liquid & 
      More layering\newline More low frequencies\newline Pitch modulation & 
      Frequency  \\
        \hline
        \textit{Drama (C)} & 
        Liquid\newline Sizzling pan\newline Droplets & 
        Emphasize high frequencies\newline Emphasize high frequencies\newline Stronger first impact & 
        Frequency \\
        \hline
        \textit{Drama/Comedy (LA)} & 
        Waves\newline Bird\newline Air & 
        More layering\newline Equalization\newline More low frequencies & 
        Temporal and frequency  \\
        \hline
      \textit{Drama/Comedy (C)} & 
      Bubbles\newline Air\newline Waves & 
      More reverberation\newline Pitch modulation\newline Stronger first impact & 
      Spectro-temporal \\
        \hline
        \textit{Sci-fi (LA)} &  
        Spaceship\newline Hits\newline Explosions & 
        Emphasize high frequencies\newline More low frequencies\newline Stronger first impact & 
        Temporal and frequency \\
        \hline
        \textit{Sci-fi (C)} & 
        Spaceship\newline Laser\newline Hits & 
        More layering\newline More reverberation\newline Stronger first impact & 
        Temporal and frequency \\
        \hline
    \end{tabular}
}
\caption{Aspects to be changed in each sound effect and the classification for optimization. C: Cartoon. LA: Live-action}
\label{Optimization}
\end{table*}

Table 4 helps to see how the use of the synthetic samples is perceived, and highlights the feature domain where sound designers might start looking to optimise their models.These initial experiments have provided valuable insights into the features that describe and distinguish realistic from synthetic sounds. For example, we know that for the synthetic model of an explosion we would need to pay attention to the frequency of the first impact to create a more believable sound, and for waves we would need to pay attention to the duration and provide more than one audio file to achieve the desired realism. 

\section{Discussion and conclusions}

Procedural audio models offer easy access and utility for sound designers. While their popularity is growing within the community, there is still a gap between industry standards and developers when it comes to optimising these models.

Our evaluation revealed different perceptions of audio realism between animated and live-action scenes. Certain synthesis models were found to be less believable, while others showed potential for optimisation. 

\begin{enumerate}
    \item \textbf{Contextual believability:} Metrics of believability vary depending on the intention and context of the audiovisual project. For example, in scenes simulating daily routines, procedural audio samples were perceived as less real than non-daily sounds. In both animated and live-action drama scenes, believability was lower than in action scenes.
 
    \item \textbf{Genre-specific performance:} Some synthetic models performed better in live-action sci-fi films than in cartoons. This suggests that genre and context have a significant impact on the perceived realism of synthesised audio.
\end{enumerate}

There is no official evaluation method for sound synthesis in audio visual projects yet, though \cite{Design} reviewed two methods used in virtual acoustic spaces, evaluating the design tools process. Sound design relies heavily on the sound designer's ability to manipulate audio samples to fit the narrative of the story. Individual perceptions and biases play a role in determining how each on-screen object should sound. Despite these differences, our study provides a valuable database and starting point for sound designers to improve their models.

Finally we could answer our main research question in 3 points 
\begin{itemize}
    \item \textbf{Perception metrics:} We found that credibility metrics depend on the context and goals of the audiovisual project. This finding addresses our main research question.
  
    \item \textbf{Performance scenarios:} We identified scenarios where synthetic samples outperformed real ones, answering our second research question.
 
    \item \textbf{Model optimisation:} Through online listening tests and feedback correlation, we identified at least three sound effects that require optimisation. This feedback highlights the audio features that need better tuning or theoretical adjustments to improve perceived realism.
\end{itemize}

Our research highlights the importance of context and genre in the perception of procedural audio samples. The findings provide a basic understanding for audio developers, guiding them on which features to optimise for enhanced realism and believability in different audiovisual contexts.

\section{Future work}
By deepening our understanding of procedural audio models, we can bridge the gap between industry standards and developer capabilities, leading to more effective and persuasive sound design in different audiovisual contexts.

Currently, sound designers face challenges in understanding and manipulating procedural audio samples. The use of machine learning and neural networks for feature importance analysis could help identify the critical features of each sound category, providing sound designers with robust tools for refinement. Further psychoacoustic testing should be carried out to understand the difference in perception between synthetic and live, between live action and cartoon.

In addition, expanding the dataset to include a wider range of sound categories will improve the insights available to developers and give sound designers confidence that procedural audio models can reliably generate believable, high-quality sounds.

 \bibliographystyle{splncs04}
\bibliography{Springer_Latex_Template/sample_saj}

\begin{thebibliography}{10}
\providecommand{\url}[1]{\texttt{#1}}
\providecommand{\urlprefix}{URL }
\providecommand{\doi}[1]{https://doi.org/#1}

\bibitem{foleygrail}
Ament, V.T.: The Foley grail: The art of performing sound for film, games, and animation. Routledge (2014)

\bibitem{AQA}
AQA: A-level drama and theatre: Sound design guidance (2020), \url{https://filestore.aqa.org.uk/resources/drama/AQA-7262-SOUND-D-TG.PDF}, accessed: 2024-07-19

\bibitem{essentia}
Bogdanov, D., Wack, N., E.~Góme~and, S.G.e.a.: Essentia: An open-source library for sound and music analysis. In: 21st ACM International Conference on Multimedia. Association for Computing Machinery (October 2013)

\bibitem{rfc}
Breiman, L.: Random forests. Machine Learning pp. 5--32 (01 2001)

\bibitem{xgboost}
Chen, T., Guestrin, C.: XGBoost: {A} Scalable Tree Boosting System, pp. 785--794 (06 2016)

\bibitem{Whoosh}
Cherny, E., Lilius, J., Mouromtsev, D.: A method for automatic whoosh sound description. In: Torin, A., Hamilton, B., Bilbao, S., Newton, M. (eds.) Proceedings of the 20th International Conference on Digital Audio Effects (DAFx-17), Edinburgh, UK, September 5–9, 2017. p. 459–465. University of Edinburgh (2017)

\bibitem{dakic}
Dakic, V.: Sound Design for Film and Television. GRIN Verlag (2009), \url{https://books.google.co.uk/books?id=wSVPl1TSOm8C}

\bibitem{ircam}
Hug, D., Misdariis, N.: Towards a conceptual framework to integrate designerly and scientific sound design methods (09 2011)

\bibitem{Design}
Liljedahl, M., Fagerl\"{o}nn, J.: Methods for sound design: a review and implications for research and practice. Association for Computing Machinery, New York, NY, USA (2010)

\bibitem{Lopez}
Lopez, M., Kearney, G., Hofstädter, K.: Seeing film through sound: Sound design, spatial audio and accessibility for visually impaired audiences. British Journal of Visual Impairment  \textbf{40} (05 2020)

\bibitem{shap}
Lundberg, S.M., Lee, S.I.: A unified approach to interpreting model predictions. 31st Conference on Neural Information Processing Systems (NIPS)  (2017)

\bibitem{hyper}
Martusciello, F.: The reality of the reproduction. aesthetics of a ”conscious” approach to sound design in the soundscape composition: a case study. Association for Computing Machinery, New York, NY, USA (2022)

\bibitem{ObjMoffat}
Moffat, D., Reiss, J.: Objective evaluations of synthesised environmental sounds. (09 2018)

\bibitem{MoffatSS}
Moffat, D., Selfridge, R., Reiss, J.: Sound Effect Synthesis, pp. 274--299 (06 2019)

\bibitem{realism}
Murray, L.: Authenticity and realism in documentary sound. The Soundtrack  \textbf{3},  131--137 (12 2010)

\bibitem{pic}
Piczak, K.J.: Esc: Dataset for environmental sound classification. p. 1015–1018. Association for Computing Machinery, New York, NY, USA (06 2015)

\bibitem{Rochesso}
Rocchesso, D., Bresin, R., Fernström, M.: Sounding objects. Multimedia, IEEE  \textbf{10},  42--52 (05 2003)

\bibitem{webmushra}
Schoeffler, M., Bartoschek, S., Stöter, F.R., Roess, M., Westphal, S., Edler, B., Herre, J.: webmushra — a comprehensive framework for web-based listening tests. Journal of Open Research Software  \textbf{6} (02 2018)

\bibitem{bonferroni}
Sedgwick, P.: Multiple significance tests: the bonferroni correction. BMJ (online)  \textbf{344} (01 2012)

\bibitem{pauletto}
Selfridge, R., Pauletto, S.: Investigating the sound design process: two case studies from radio and film production. In: Design Research Society, DRS Digital Library (2022)

\bibitem{PCA}
Shlens, J.: A tutorial on principal component analysis. CoRR-Computer science bibliography  (2014)

\bibitem{susini}
Susini, P., Houix, O., Misdariis, N.: Sound design: An applied, experimental framework to study the perception of everyday sounds. The New Soundtrack  \textbf{4},  103--121 (09 2014)

\bibitem{cuidado}
Vinet, H., Herrera, P., Pachet, F.: The cuidado project. (01 2002)

\bibitem{low}
Winters, P.: Sound design for low and no budget films. Routledge (2017)

\bibitem{interp}
Zinemanas, P., Rocamora, M., Miron, M., Font, F., Serra, X.: An interpretable deep learning model for automatic sound classification. Electronics  \textbf{10}, ~850 (04 2021)

\end{thebibliography}

\end{document}